\documentclass[pre,onecolumn,showpacs,preprintnumbers,amsmath,amssymb]{revtex4}
\usepackage{graphics}

\begin{document}

\title{Electronic damping of molecular motion at metal surfaces}
\author{J. R. Trail}\email{jrt32@cam.ac.uk}
\author{M. C. Graham}
\author{D. M. Bird}
\affiliation{Department of Physics, University of Bath, Bath BA2 7AY, UK}

\date{January, 2001}

\begin{abstract}
A method for the calculation of the damping rate due to electron-hole pair excitation for atomic and molecular motion at metal surfaces is presented.
The theoretical basis is provided by Time Dependent Density Functional Theory (TDDFT) in the quasi-static limit and calculations are performed within a standard plane-wave, pseudopotential framework.
The artificial periodicity introduced by using a super-cell geometry is removed to derive results for the motion of an isolated atom or molecule, rather than for the coherent motion of an ordered over-layer.
The algorithm is implemented in parallel, distributed across both ${\bf k}$ and ${\bf g}$ space, and in a form compatible with the CASTEP code.
Test results for the damping of the motion of hydrogen atoms above the Cu(111) surface are presented.
\end{abstract}

% keywords: electron-hole pair creation; friction; molecule-surface dynamics
% Interaction of atoms, molecules and their ions with surfaces; photon and electron emission; nuetralisation of ions
% Atomic, molecular and ion beam impact and interactions with surfaces
% Surface enhanced molecular states and other gas-surface interactions
% plane-wave methods

\maketitle

\section{Introduction}

Considerable progress has been made in recent years in understanding the fundamental processes involved in gas-surface interactions.
This has been based on the parallel developments of large-scale electronic structure calculations based on density functional theory, combined with multi-dimensional quantum and classical analysis of the dynamics \cite{bird97}.
Despite these advances there remains one key area that is still largely unexplored and poorly understood; the process of energy dissipation into substrate degrees of freedom.
Although this is known to be of central importance in many situations \cite{darling95}, there exist no `real' calculations to date for the energy loss to either phonons or electrons in the surface. 

In particular there have been a number of recent experiments that have shown convincing evidence that energy dissipation by the creation of electron-hole pairs is a significant effect in gas-surface dynamics.
Gostein et al \cite{gostein97} carried out a detailed state-to-state analysis of H$_2$ scattering from Pd(111) and showed that, for example, in the vibrational relaxation of ($\nu=1,J=1$) to ($\nu=0,J=5$) an average of 120 meV is lost to the substrate during the scattering event, presumably to electron-hole pair formation.
Nienhaus and co-workers \cite{nienhaus99} measured directly the hot electrons and holes created at Ag and Cu surfaces by the adsorption of thermal hydrogen and deuterium in the form of `chemicurrents' in a Schottky diode.
Finally, Huang et al \cite{huang00} have studied NO scattering from Au(111) and have concluded that the main sink of energy for the vibrational relaxation of $\nu=2$ molecules is the surface, with the strong dependence of the de-excitation probability on incident energy providing evidence that an electron-hole pair mechanism is the dominant factor.

We carry out a calculation of the ground state properties of an interacting surface/molecule system using a plane-wave basis and a super-cell geometry, and use these results to evaluate the friction coefficient associated with the motion of a molecule at a chosen position and in a direction of choice.
This is achieved using the well established `Golden Rule' expression \cite{grimvall81,hellsing84,liebsch97} that may be obtained by applying Time Dependent Density Functional Theory (TDDFT) together with a quasi-static limit \cite{hellsing84}, or less stringently by applying the Golden Rule directly to the available Kohn-Sham states \cite{grimvall81}.
Essentially the theory is as described by Hellsing and Persson \cite{hellsing84} or Liebsch \cite{liebsch97}.
The super-cell method has the advantage that it retains the continuous spectrum of one-electron excitations, unlike cluster models \cite{headgordon92}, and this is important for the interactions considered here.
In first principles calculations of molecule-surface systems a super-cell of sufficient size is usually chosen to prevent any significant interaction between the adsorbates in neighbouring super-cells.
When considering electron-hole pair excitation a slightly more subtle effect must be taken into account, arising from the enforced periodicity of the perturbation that produces the electron-hole pairs.
We are primarily interested in the energy loss by electron-hole pair excitation due to the motion of an isolated molecule interacting with the surface, whereas the super-cell geometry will naturally describe the damping of an ordered over-layer.
For the periodic system, conservation of crystal momentum prevents transitions occurring that will occur for an isolated molecule interacting with the surface.
Results for an isolated molecule are derived from the available periodic perturbation, and a significant difference is found between the energy loss behaviour of the periodic and isolated systems.

To test our method we investigate the friction coefficient of an H atom above the hcp hollow site of a Cu(111) surface.
Spin is included explicitly in the Kohn-Sham theory using the gradient corrected local spin density approximation for exchange-correlation (LSDA-GC).
In the next section the evaluation of the dynamic self-energy and friction coefficient from Kohn-Sham results using a plane-wave basis is described.
In section \ref{sec:alg} the implementation of this as a parallel algorithm is described, along with a brief description of the performance of the algorithm.
Results for H/Cu(111) are discussed in section \ref{sec:results}, and section \ref{sec:conc} is the conclusion.

\section{Theory}
\label{sec:theory}
The experimentally measurable energy loss spectrum for a particular mode is directly related to the dynamic self-energy, $\Lambda(\omega)$.
Since we are interested in the energy loss, the imaginary part of this self-energy is the required quantity and this can be expressed using \cite{hellsing84}
\begin{eqnarray}
\mathrm{Im}\ \Lambda(\omega) =
            \frac{2 \pi e^2}{M}
            \sum_{{\bf k},{\bf k}'} \sum_{n,n'}
            \left| \langle 
            \psi_{\bf k}^n | \phi^{\mathrm{eff}}({\bf r},\omega) |\psi_{{\bf k}'}^{n'}
            \rangle \right| ^2
            \nonumber\\ \times 
            \left( f(\epsilon_{\bf k}^n) - f(\epsilon_{{\bf k}'}^{n'}) \right)
            \delta( \hbar\omega-\epsilon_{\bf k}^n+\epsilon_{{\bf k}'}^{n'} )
\label{e2.1}
\end{eqnarray}
where $\psi_{\bf k}^n$ and $\epsilon_{\bf k}^n$ are the Kohn-Sham wavefunctions and energies resulting from a density-functional description of the ground state, and $f(\epsilon)$ is the Fermi-Dirac occupation function.
Spin degeneracy is assumed in Eq. (\ref{e2.1}), hence the factor of 2; the extension to spin polarised systems is straightforward.
The effective field, $\phi^{\mathrm{eff}}$, is the TDDFT effective field with contributions from the field of the displaced nuclei, the Coulomb field of the induced charge density and a contribution from exchange-correlation.
Equation (\ref{e2.1}) can be derived by direct application of the Golden-Rule and a single electron approximation \cite{grimvall81}, or through TDDFT with the assumption of a slow time-dependent perturbation (see \cite{hellsing84} and \cite{amusia98} for more clarification of the role of TDDFT).

Equation (\ref{e2.1}) provides the rate of energy loss of a mode of frequency $\omega$ due to the excitation of electron-hole pairs as $\tau^{-1}= \mathrm{Im}\ \Lambda(\omega) / \omega$.
Taking the quasi-static limit $\omega \rightarrow 0$ results in the rate of energy loss for the motion of the atom, and this can be expressed as the friction coefficient $\eta$ defined via the standard Langevin equation.
For motion in the direction $\hat{\mathbf{h}}$, $\eta$ is given by (see Hellsing and Persson \cite{hellsing84}, or Plihal and Langreth \cite{plihal98})
\begin{eqnarray}
\eta =& M \lim_{\omega \rightarrow 0}  \mathrm{Im}\ \Lambda(\omega)/\omega
         \nonumber\\
       =& 2 \pi \hbar
         \sum_{{\bf k},{\bf k}'} \sum_{n,n'}
         \left| \langle 
         \psi_{\bf k}^n | \hat{\mathbf{h}}.\nabla V |\psi_{{\bf k}'}^{n'}
         \rangle \right| ^2
         \delta( \epsilon_{\mathrm{F}} - \epsilon_{{\bf k} }^{n } )
         \delta( \epsilon_{\mathrm{F}} - \epsilon_{{\bf k}'}^{n'} )
\label{e2.2}
\end{eqnarray}
where $\epsilon_{\mathrm{F}}$ is the Fermi energy, $M$ is the mass of the molecule, and $\hat{\mathbf{h}}.\nabla V$ is the static limit of $-e\phi^{\mathrm{eff}}$ given by the derivative of the Kohn-Sham self-consistent potential in the direction $\hat{\mathbf{h}}$.

Equation (\ref{e2.2}) is evaluated using wavefunctions of the form
\begin{equation}
\psi_{\bf k}^n ({\bf r}) 
       = u^n_{\bf k} \mathrm{e}^{ \mathrm{i}{\bf k}.{\bf r} }
       =\frac{1}{\sqrt{V_0}} \sum_{\bf g} 
            C^n_{\bf g}({\bf k}) \mathrm{e}^{ \mathrm{i}({\bf k}+{\bf g}).{\bf r} }
\label{e2.4}
\end{equation}
where $V_0$ is the volume of the super-cell, and only the wavefunctions sampled at a finite number of ${\bf k}$ points in the irreducible wedge of the Brillouin Zone are available.
The change in the Kohn-Sham potential due to the motion of an isolated atom is not directly available from a super-cell calculation, but we can obtain the change due the motion of a periodic lattice of atoms, $\delta V^{\mathrm{KS}}_{\mathrm{lattice}}$, as a finite difference.

\subsection{Obtaining the interaction of an isolated molecule from that of an over-layer}

To calculate the dynamic interaction of an isolated molecule with a surface within a super-cell geometry care must be taken with the interpretation of Eq. (\ref{e2.1}).
If $\hat{\mathbf{h}}.\nabla V$ is obtained directly as a periodic function, drastic consequences result - the integral becomes zero for ${\bf k} \neq {\bf k}'$.
This is due to the super-cell geometry enforcing a conservation of crystal momentum that would be correct for describing the interaction of a real over-layer of molecules in coherent motion, but is not physically realistic for the aperiodic single molecule/surface system that concerns us here.
In light of this we must obtain the change in the Kohn-Sham potential due to the motion of an isolated molecule, $\delta V^{\mathrm{KS}}_{\mathrm{isolated}}$, from the change due to the coherent motion of an over-layer, $\delta V^{\mathrm{KS}}_{\mathrm{lattice}}$.
The relationship between the two can be found by applying linear response theory \cite{gross85}.
Although the theory is given here for a local pseudopotential, the generalisation to non-local pseudopotentials is straightforward.

We begin by considering the change in the total pseudopotential, $\delta \phi ( {\bf r} )$, due to an infinitesimal change in the position of the atom in each super-cell, $\delta {\bf u}_{\bf x}( {\bf l} )$.
With the pseudopotential due to an atom at ${\bf x}$ denoted $V_{\mathrm{pseudo}}({\bf r} - {\bf x})$ this takes the form
\begin{equation}
\delta \phi ( {\bf r} ) = \sum_{\bf l}
                 \nabla V_{\mathrm{pseudo}}({\bf r} - {\bf x} - {\bf l}).
                 \delta {\bf u}_{\bf x}( {\bf l} ),
\end{equation}
where  ${\bf l}$ is a lattice vector.
The change in the Kohn-Sham potential can be obtained from $\delta \phi$ via the static inverse dielectric function, $\epsilon^{-1}_{\mathrm{KS}}({\bf r},{\bf r}')$ that corresponds to the original Kohn-Sham calculation \cite{inkson86}.
This gives
\begin{equation}
\delta V^{\mathrm{KS}}_{\mathrm{lattice}} ({\bf r}) = \int \epsilon^{-1}_{\mathrm{KS}} ({\bf r},{\bf r}') \delta \phi( {\bf r}' ) \mathrm{d}^3{\bf r}'
\end{equation}
or
\begin{equation}
\delta V^{\mathrm{KS}}_{\mathrm{lattice}} ({\bf r}) = \sum_{\bf l} {\bf R}_{\bf x} ( {\bf r},{\bf l} ).\delta {\bf u}_{\bf x}( {\bf l} )
\end{equation}
where
\begin{equation}
{\bf R}_{\bf x} ( {\bf r},{\bf l} ) =  \int \epsilon^{-1}_{\mathrm{KS}} ({\bf r},{\bf r}') \nabla V_{\mathrm{pseudo}}({\bf r}' - {\bf x} - {\bf l}) \mathrm{d}^3{\bf r}'
\end{equation}
is the derivative of the Kohn-Sham potential with respect to the change in the position of an atom at ${\bf x}+{\bf l}$.

For an isolated atom we require $\delta V^{\mathrm{KS}}_{\mathrm{isolated}} ({\bf r})$, the change in the Kohn-Sham potential due to the change in the position of an atom at ${\bf x}$.
The formally correct way to obtain this is to obtain $\epsilon^{-1}_{\mathrm{KS}} ({\bf r},{\bf r}')$ from the original DFT calculation (for an example of this type of calculation see Godby et al \cite{godby88}), obtain ${\bf R}_{\bf x} ( {\bf r},{\bf l} )$, and use this to evaluate
\begin{equation}
\delta V^{\mathrm{KS}}_{\mathrm{isolated}} ({\bf r}) = {\bf R}_{\bf x} ( {\bf r},{\bf 0} ).\delta {\bf u}_{\bf x}( {\bf 0} ).
\end{equation}
However, provided ${\bf R}_{\bf x} ( {\bf r},{\bf l} )$ is well localised around the site of the perturbed atom such that there is little overlap between the responses to the motion of atoms in adjacent unit cells we may take 
\begin{eqnarray}
\delta V^{\mathrm{KS}}_{\mathrm{isolated}}({\bf r}) & \approx & \Theta( {\bf r} ) \sum_{\bf l} {\bf R}_{\bf x} ( {\bf r},{\bf l} ).\delta {\bf u}_{\bf x}( {\bf l} ) \\ \nonumber
                                  & \approx  & \Theta( {\bf r} ) \delta V^{\mathrm{KS}}_{\mathrm{lattice}} ({\bf r})
\end{eqnarray}
where $\Theta( {\bf r} )=1$ within a Wigner-Seitz cell centred on the perturbed atom, and is zero elsewhere.

From the super-cell calculation the Kohn-Sham potential for the atom at ${\bf x}\pm{\bf h}$ is evaluated to obtain $\hat{\mathbf{h}}.\nabla V$ as
\begin{eqnarray}
\hat{\mathbf{h}}.\nabla V & = &
\frac{\delta V^{\mathrm{KS}}_{\mathrm{isolated}}}{\delta |{\bf u_{\bf x}}({\bf 0})|} \nonumber \\
 & \approx &  
\frac{ \Theta (\mathbf{r}) }{2|{\bf h}|}
\left( 
        [V^{\mathrm{KS}}_{\mathrm{lattice}} - \epsilon_{\mathrm{F}}]_{{\bf x}+{\bf h}}-
        [V^{\mathrm{KS}}_{\mathrm{lattice}} - \epsilon_{\mathrm{F}}]_{{\bf x}-{\bf h}}
\right)
\label{e2.3}
\end{eqnarray}
where the variation of the zero of the Kohn-Sham potential has been corrected for using the Fermi energy associated with the self-consistent results for each atomic position.

Equation (\ref{e2.3}) will be accurate providing the response is isolated within the Wigner-Seitz cell centred on the perturbed atom (or, equivalently, $V^{\mathrm{KS}}_{\mathrm{lattice}} ({\bf r})$ is negligible at the border of this Wigner-Seitz cell), and $|{\bf h}|$ is small enough.
In practice Eq. (\ref{e2.3}) corresponds to reducing the volume of integration in Eq. (\ref{e2.1}) from the entire lattice to one Wigner-Seitz cell.
In section \ref{sec:results} the consequences of considering the motion of an isolated atom, as described above, are investigated by comparing results with those obtained by treating the motion as that of an ordered over-layer of atoms.

\subsection{Evaluation for a plane-wave basis}

A plane-wave calculation results in a set of Kohn-Sham states on a finite grid (in real space or reciprocal space) sampled at a finite number of ${\bf k}$ points and for a finite number of bands, hence we must obtain the discrete equivalent of Eq. (\ref{e2.2}) to evaluate $\eta$.
It is also desirable to apply the space-group symmetry to reduce the number of ${\bf k}$ points that must be considered.
This is achieved as follows.

First Eq. (\ref{e2.1}) is discretised using a uniform grid of ${\bf k}$ points throughout the $1^{\mathrm{st}}$ Brillouin Zone, and a conventional smearing function
\begin{eqnarray}
\mathrm{Im}\ \Lambda(\omega) =
            \frac{2 \pi e^2}{M} \frac{1}{N^2}
            \sum_{{\bf k},{\bf k}'} \sum_{n,n'}
            \left| \langle 
            \psi_{\bf k}^n | \phi^{\mathrm{eff}}({\bf r},\omega) |\psi_{{\bf k}'}^{n'}
            \rangle \right| ^2
            \nonumber\\ \times 
            \left( f(\epsilon_{\bf k}^n) - f(\epsilon_{{\bf k}'}^{n'}) \right)
            \tilde \delta( \hbar\omega-\epsilon_{\bf k}^n+\epsilon_{{\bf k}'}^{n'} )
\label{e2.6}
\end{eqnarray}
where $N$ is the number of ${\bf k}$ points, $f(\epsilon)$ is the Fermi-Dirac occupation function, and $-\tilde{\delta}(x)$ is the derivative of the `squashed Fermi-Dirac' \cite{white96} function used for occupation numbers in the original Kohn-Sham calculation.
$\tilde{\delta}(x)$ is given by
\begin{eqnarray}
\tilde{\delta}(x) = \frac{1}{\Delta} \frac{y}{t} \exp\left(\frac{1}{2}-y^2 \right), &\  y  = \frac{1}{\Delta} \frac{|x|}{t} + \sqrt{\frac{1}{2}}
\label{e2.7}
\end{eqnarray}
where $\Delta$ is an energy parameter describing the width of the function, and $t=\sqrt{\pi/2}$.

To obtain $\eta$ we take $\lim_{\omega \rightarrow 0} \mathrm{Im}\ \Lambda(\omega)/\omega$, resulting in
\begin{equation}
\eta = 2 \pi \hbar \frac{1}{N^2}
         \sum_{{\bf k},{\bf k}'} \sum_{n,n'}
         \left| \langle 
         \psi_{\bf k}^n | \hat{\mathbf{h}}.\nabla V |\psi_{{\bf k}'}^{n'}
         \rangle \right| ^2
         \left( f(\epsilon_{\bf k}^n) - f(\epsilon_{{\bf k}'}^{n'}) \right)
         \tilde \delta'( \epsilon_{\bf k}^n-\epsilon_{{\bf k}'}^{n'} )
\label{e2.7b}
\end{equation}
where $-\tilde \delta'(x)$ is the $2^{\mathrm{nd}}$ derivative of the `squashed Fermi-Dirac' function, and we have taken the static limit of $\phi^{\mathrm{eff}}$.
Equation (\ref{e2.7b}) reduces to Eq. (\ref{e2.2}) in the zero temperature and continuum limit.
It is important to note that Eq. (\ref{e2.2}) cannot be directly discretised by replacing the $\delta(x)$ functions with $\tilde \delta(x)$ as this corresponds to taking the continuum and zero temperature limits before discretisation.
This would result in the inclusion of a contribution to $\eta$ from transitions that should not be present (such as from a state to itself).

Next we reduce the ${\bf k}$ points that must be considered to those within the irreducible wedge of the Brillouin Zone using the space-group of the lattice.
Since the set of ${\bf k}$ points possess point group symmetry Eq. (\ref{e2.7b}) immediately takes the form
\begin{eqnarray}
\eta = 2 \pi \hbar \frac{1}{N_S^2}
         \sum_{S,S'} \sum_{{\bf k},{\bf k}'}^{\mathrm{IW}} \sum_{n,n'}
         \left| \langle 
         \psi_{P{\bf k}}^n | \hat{\mathbf{h}}.\nabla V |\psi_{P'{\bf k}'}^{n'}
         \rangle \right| ^2
         \left( f(\epsilon_{\bf k}^n) - f(\epsilon_{{\bf k}'}^{n'}) \right)
         \nonumber\\ \times
         \tilde \delta'( \epsilon_{\bf k}^n-\epsilon_{{\bf k}'}^{n'} )
         w_{{\bf k} } w_{{\bf k}'}
\label{e2.8}
\end{eqnarray}
where $\mathrm{IW}$ denotes a sum over points in the irreducible wedge, and $S$,$S'$ are space-group operators composed of a unitary transformation, $P$, and a non-symmorphic translation, ${\bf w}$.
The number of operators in the space-group is denoted by $N_S$, and $w_{{\bf k} } = N_{{\bf k} }/N$ where $N_{{\bf k} }$ is the number of distinct ${\bf k}$ points in the entire Brillouin Zone created by applying the complete set of $N_S$ space group operators to the point ${\bf k}$ in the irreducible wedge.
There are no symmetry operators associated with the eigenvalues since $\epsilon_{{\bf k}}^n = \epsilon_{P{\bf k}}^n$.
Transforming the integrand in real space by $S'^{-1}$ and using the identity \cite{altmann91}
\begin{equation}
S \psi_{{\bf k}}^n({\bf r}) = \psi_{{\bf k}}^n(P{\bf r} + {\bf w} )
                            = \psi_{P{\bf k}}^n({\bf r})
\label{e2.9}
\end{equation}
results in
\begin{eqnarray}
\eta = 2 \pi \hbar \frac{1}{N_S}
         \sum_{S} \sum_{{\bf k},{\bf k}'}^{\mathrm{IW}} \sum_{n,n'}
         \left| \langle
         \psi_{P{\bf k}}^n | \hat{\mathbf{h}}.\nabla V |\psi_{{\bf k}'}^{n'}
         \rangle \right| ^2
         \left( f(\epsilon_{\bf k}^n) - f(\epsilon_{{\bf k}'}^{n'}) \right)
         \nonumber\\ \times
         \tilde \delta'( \epsilon_{\bf k}^n-\epsilon_{{\bf k}'}^{n'} )
         w_{{\bf k} } w_{{\bf k}'}
\label{e2.10}
\end{eqnarray}
where repeated sums over the same operator are removed and $\hat{\mathbf{h}}.\nabla V$ is required to possess the space-group symmetry corresponding to the available ${\bf k}$ points.
This requirement means that the original Kohn-Sham calculations must be performed with symmetry low enough to allow the motion of the atom in the direction we are interested in, even if the instantaneous position of the atom corresponds to a higher symmetry.
 In practise we also only calculate the quantities within the sum where the $\left( f(\epsilon_{\bf k}^n) - f(\epsilon_{{\bf k}'}^{n'}) \right)\tilde \delta'( \epsilon_{\bf k}^n-\epsilon_{{\bf k}'}^{n'} )$ factor is  greater than some small value, preventing the cost of calculating matrix elements that make negligible contribution to $\eta$ and limiting the sum to a finite number of bands.

For the the motion of an over-layer of atoms, where $\hat{\mathbf{h}}.\nabla V$ is periodic, a similar treatment results in
\begin{equation}
\eta_{ \mathrm{\bf k} = \mathrm{\bf k}' } = 2 \pi \hbar
         \sum_{{\bf k}}^{\mathrm{IW}} \sum_{n,n'}
         \left| \langle
         \psi_{ {\bf k}}^n | \hat{\mathbf{h}}.\nabla V |\psi_{{\bf k} }^{n'}
         \rangle \right| ^2
         \left( f(\epsilon_{\bf k}^n) - f(\epsilon_{{\bf k}}^{n'}) \right)
         \tilde \delta'( \epsilon_{\bf k}^n-\epsilon_{{\bf k}}^{n'} )
         w_{{\bf k} }
\label{e2.10b}
\end{equation}
where only ${\bf k} = {\bf k}'$ transitions contribute.

\begin{figure}
\begin{center}
\includegraphics*{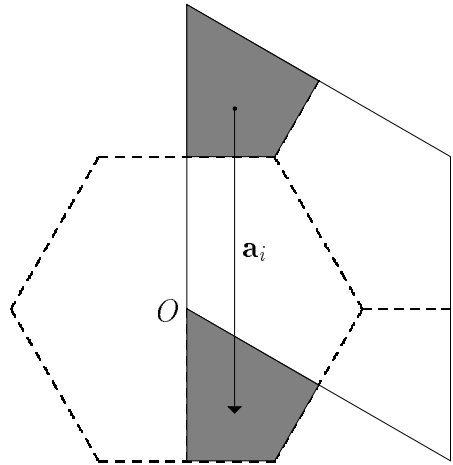}
\end{center}
\caption{For an atom at $O$ integration must be performed over the Wigner-Seitz unit cell centred at $O$ (dashed line).
For the parallelogram (which represents the originally chosen unit cell) the function $\hat{\mathbf{h}}.\nabla V$ is not localised within the unit cell, and does not possess space-group symmetry.}
\label{fig1}
\end{figure}

In order to obtain the matrix elements we require the Kohn-Sham states corresponding to the images of the available ${\bf k}$ points under transformation by a point group operator, $\psi_{P{\bf k}}^n$.
Transforming $\psi_{{\bf k}}^n$ in reciprocal space gives us the coefficients of each $\psi_{P{\bf k}}^n$, and this is achieved by applying Eq. (\ref{e2.9}) to Eq. (\ref{e2.4}) and projecting out the required coefficients to give
\begin{equation}
C^n_{\bf g}(P{\bf k}) = C^n_{P^{-1}{\bf g}}( {\bf k}) \mathrm{e}^{-\mathrm{i}( P{\bf k} +{\bf g}).{\bf w}  }.
\label{e2.12}
\end{equation}
The wavefunctions in real space are then obtained using a FFT
\begin{equation}
u^{n }_{ {\bf k} } = \mathrm{FFT} [ C^n_{\bf g}( {\bf k}) ]
\end{equation}
 and the required integrals are evaluated as sums over a unit cell (this is analytically correct for a plane-wave basis).
This cell must be chosen such that $\hat{\mathbf{h}}.\nabla V$ is negligible at its borders and such that the `truncated' potential possesses the correct space-group symmetry.
As mentioned previously the ideal choice is a Wigner-Seitz cell centred at the interacting atom, as is shown schematically in Fig.\ \ref{fig1} for a $2d$ hexagonal unit cell with the atom of interest at $O$.
To perform the sum it is necessary to divide the integration volume into different regions $X_i$ such that transforming region $X_i$ by a vector ${\bf a}_i$ constructs the Wigner-Seitz cell from the original unit cell.
The matrix element then takes the form
\begin{equation}
\langle
\psi_{P{\bf k}}^n | \hat{\mathbf{h}}.\nabla V |\psi_{{\bf k}'}^{n'}
\rangle = \
\sum_{i}
\frac{1}{N_{\bf r}}  \sum_{ j }
u^{n*}_{P{\bf k}}( {\mathbf{r}}_j )
\hat{\mathbf{h}}.\nabla V( {\mathbf{r}}_j )
u^{n'}_{{\bf k}'} ( {\mathbf{r}}_j )
\mathrm{e}^{\mathrm{i}({\bf k}'-P{\bf k}).({\bf r}_j + {\bf a}_i) }
\label{e2.13}
\end{equation}
where $\hat{\mathbf{h}}.\nabla V$ and the wavefunctions are available over a grid of $N_{\bf r}$ real space points, ${\bf r}_j$.

It should be noted that for an isolated atom the integral is carried out over one unit cell as a consequence of $\hat{\mathbf{h}}.\nabla V $ being localised in that cell, whereas for the coherent motion of an over-layer the integral is carried out over one unit cell as a consequence of the periodicity of the integrand.
Hence, to evaluate $\eta_{\mathrm{\bf k}=\mathrm{\bf k}'}$ the rearrangement of Eq. (\ref{e2.13}) is not necessary.

\section{Implementation as a parallel algorithm}
\label{sec:alg}
We are interested in large scale systems, hence a parallel implementation of both the original Kohn-Sham calculation and the evaluation of $\eta$ is desirable.
Multiprocessor algorithms for plane-wave Kohn-Sham methods are available, so we describe the latter only.

\begin{figure}
\begin{center}
\includegraphics*{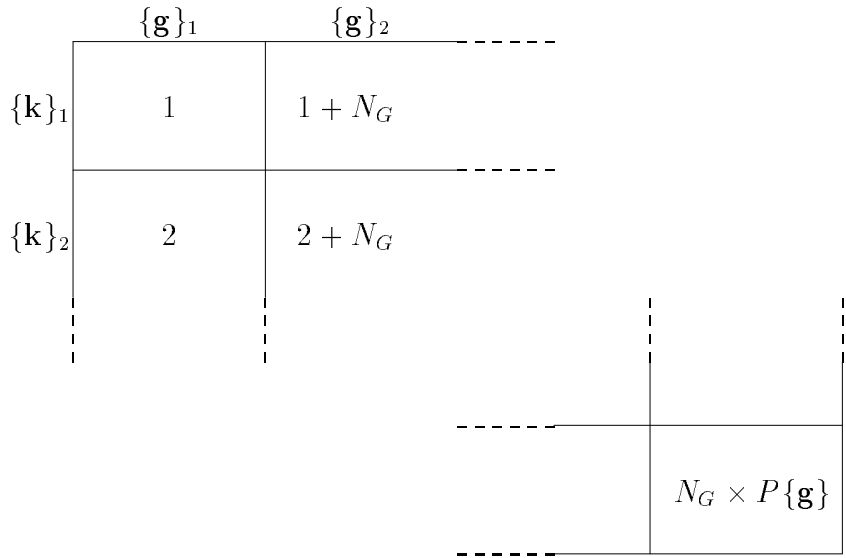}
\end{center}
\caption{Distribution of data across processors - each box is one processor and contains the processor number.
Columns are processors that hold data for subregions of real or reciprocal space, and rows are processors that hold data for subsets of ${\bf k}$ points.}
\label{fig2}
\end{figure}

First we consider the distribution of data between the $N_p$ available processors.
These are divided into $N_{G}$ groups of $P\{ {\bf g} \}$ processors, and the ${\bf k}$ points are distributed equally between these groups of processors.
Data specific to a ${\bf k}$ point is stored on the associated group.
The ${\bf g}$ points are then distributed across the $P\{ {\bf g} \}$ processors in each group, and data associated with each ${\bf g}$ point held on the associated processor.
Data that is not dependent on ${\bf k}$ is the same for each group but distributed across processors within the group, and data that is not dependent on ${\bf g}$ is the same within each group but distributed across groups.
This distribution is shown figuratively in Fig.\ \ref{fig2}, where the ${\bf k}$ points in the $n^{\mathrm{th}}$ group are denoted $\{ {\bf k} \}_n$, and the ${\bf g}$ vectors on the $n^{\mathrm{th}}$ processor in each group are denoted $\{ {\bf g} \}_n$.

Wavefunctions in real space are obtained by a parallel FFT, so real space points are distributed as for reciprocal space points, and this FFT is performed in parallel across all processors in each group.
As efficient parallel FFT algorithms are available and a distributed sum is trivial these sections of the algorithm parallelise well.
The integral must also be must be summed over {\em pairs\/} of ${\bf k}$ points, with each point in a different group, and it is this part of the algorithm that requires a large amount of waiting and communication, and is not as efficient.
The algorithm for evaluation of Eq. (\ref{e2.10}) takes the following form, with the current processor in group $i$.

\newcommand{\keyw}[1]{{\bf #1}}
\begin{tabbing}[t]

\keyw{for} \= $n:=1,N_{\mathrm{bands}}$                \\
 \> \keyw{for} \= ${\bf k} \in \{ {\bf k} \}_i$        \\
 \> \> \keyw{for} \= $S \in $ Space-group              \\
                                    \\
 \> \> \> $C_{\bf g}^n(P{\bf k}) := $\keyw{Transform}[$C_{\bf g}^n({\bf k})$] \\
 \> \> \> $u_{P{\bf k}}^n({\bf r}) := \mathrm{FFT}[C_{\bf g}^n(P{\bf k})]$              \\
                                    \\
 \> \> \> \keyw{for} \= $n':=1,N_{\mathrm{bands}}$                 \\
 \> \> \> \> \keyw{for} \= $m:=1,N_G$                              \\
 \> \> \> \> \> \keyw{for} \= ${\bf k}' \in \{ {\bf k} \}_i$       \\
                                    \\
 \> \> \> \> \> \> \keyw{If} $m = i$ \keyw{then} $u_{{\bf k}'}^{n'}({\bf r}) := \mathrm{FFT}[ C_{\bf g}^{n'}({\bf k}') ]$ \\
 \> \> \> \> \> \> \keyw{Bcast} $u_{{\bf k}'}^{n'}({\bf r})$, ${\bf k}'$, $w_{{\bf k}'}$, $\epsilon_{{\bf k}'}^{n'}$ from group $m$ to all groups  \\
 \> \> \> \> \> \> Sum $ u_{P{\bf k}}^{n*}({\bf r})\ \hat{\mathbf{h}}.\nabla V\ u_{{\bf k}'}^{n'}({\bf r}) \mathrm{e}^{\mathrm{i}({\bf k}'-P{\bf k}).({\bf r} + {\bf a}) }$ over group $i$ \\
 \> \> \> \> \> \> Evaluate $\left( f(\epsilon_{\bf k}^n) - f(\epsilon_{{\bf k}'}^{n'}) \right) \tilde \delta'( \epsilon_{\bf k}^n-\epsilon_{{\bf k}'}^{n'} )$ \\
 \> \> \> \> \> \> Add contribution to $\eta$ using weights \\
 \> \> \> \> \> \> \ \ $w_{{\bf k}}$,$w_{{\bf k}'}$, $\left( f(\epsilon_{\bf k}^n) - f(\epsilon_{{\bf k}'}^{n'}) \right) \tilde \delta'( \epsilon_{\bf k}^n-\epsilon_{{\bf k}'}^{n'} )$  \\
                                    \\
 \> \> \> \> \> \keyw{end}\\
 \> \> \> \> \keyw{end} \\
 \> \> \> \keyw{end} \\
 \> \> \keyw{end} \\
 \> \keyw{end} \\
 \keyw{end} \\
Sum $\eta$ over all groups \\
\end{tabbing}

The subprogram \keyw{Transform:} obtains the image of the wavefunction in reciprocal space, as defined by Eq. (\ref{e2.12}).
It takes the following form, where the current processor is in the group associated with ${\bf k}$, and holds vectors $\{ {\bf g} \}_i$

\begin{tabbing}
\keyw{Transform:} \\
\keyw{for} \= $m:= 1,P\{ {\bf g} \}$ \\
  \>   \keyw{Bcast}  $\{ {\bf g}$,$ C_{{\bf g}}^n({\bf k}) \}_{m}$ from $m^{\mathrm{th}}$ to all processors in group \\
 \>   \keyw{for} \= ${\bf g}$ $\in$ $\{ {\bf g} \}_i$ \\
 \>   \> \keyw{If} $P^{-1}{\bf g}$ $\in$ $\{ {\bf g} \}_m$ \keyw{then} $C^n_{\bf g}(P{\bf k}) := C^n_{P^{-1}{\bf g}}( {\bf k}) e^{-i( P{\bf k} +{\bf g}).{\bf w}  }$ \\
 \>   \keyw{end} \\
\keyw{end} \\
\keyw{return} $C^n_{\bf g}(P{\bf k})$ \\
\end{tabbing}

It is useful to know the scaling of this algorithm with respect to the distribution of data across the processors, $(N_G,P\{{\bf g}\})$, and the number of processors available $N_p=N_G \times P\{{\bf g}\}$.
The time taken for a reasonably large system is dominated by the time spent carrying out the FFT contained within the inner loop (or waiting for another group to broadcast the result of this FFT) and is given by
\begin{equation}
t_{\parallel} \propto N_s N_{bands}^2 N_{{\bf k}}^2 \frac{1}{N_p}  P\{ {\bf g} \} t_{\mathrm{FFT}}(P\{ {\bf g} \})
\label{e3.1}
\end{equation}
where $t_{\mathrm{FFT}}(M)$ is the time taken to carry out a parallel FFT on M processors.
The speedup of the parallel FFT with increasing number of processors is complex and depends on the architecture of the parallel system \cite{gupta93}, but some general conclusions about how this effects the performance of the entire algorithm can be made.
If a linear speedup of the FFT with respect to the available processors ($P \{ {\bf g} \}$) occured then the total run time would be independent of the distribution of the processors among ${\bf k}$ and ${\bf g}$ points.
However, the actual speedup will be worse than linear due to the communication times, so for a given $N_p$ the best efficiency is achieved when the ${\bf k}$ points are distributed amongst as many processors as possible, or $P\{{\bf g}\}$ is as small as possible.

\section{Results}
\label{sec:results}

Test calculations have been carried out for a H atom moving above the hcp hollow site of a Cu(111) surface.
This system has been chosen for its simplicity (although we note that spin polarisation is needed to obtain the correct electronic structure at larger atom-surface separations) and because of its relevance to the chemicurrent experiment of Nienhaus et al \cite{nienhaus99}.
The surface is modelled by a five-layer slab, with a vacuum gap equivalent to another five layers.
A $2\times2$ in-plane super-cell is used - tests show that the deformation potential caused by the displacement of H atoms is well localised within this unit cell.
A spin-polarised version of the PW91 functional is used for exchange-correlation effects (\cite{perdew96}, and references therein), a Troullier-Martins \cite{troullier91} pseudopotential is used for Cu, and H is represented by a Coulomb potential. 
The plane-wave, pseudopotential code CASTEP is used to obtain the self-consistent potentials and Kohn-Sham states that are required for the calculation of the matrix elements in Eq. (\ref{e2.10}).
Calculations are performed with a plane-wave cut-off of 830 eV, 54 k-points are included in the full surface Brillouin zone, and a Fermi surface broadening of 0.25 eV is used.
In order to test the convergence of the method we present results for the H atom 2.5 \AA\ above the surface, and with $\hat{\mathbf{h}}$ perpendicular to the surface.

To obtain the perturbative field, $\hat{\mathbf{h}}.\nabla V$, calculations are performed with H at ${\bf x} \pm {\bf h}$.
Equation (\ref{e2.3}) is then applied to the resulting self-consistent potentials to obtain $\hat{\mathbf{h}}.\nabla V$.
The finite difference introduces two errors in the final result.
First a quadratic error is introduced by the finite difference itself, and second any small errors in the Kohn-Sham potentials are magnified for small $|{\bf h}|$.
It follows that $|{\bf h}|$ must be carefully chosen to be small enough to minimise the first of these errors, but large enough to minimise the second.
Tests for a number of $|{\bf h}|$ suggest that $|{\bf h}|=0.02$ \AA\ results in a quadratic error $\approx 0.1 \%$ and an error due to noisy Kohn-Sham potentials of $\approx 1.0 \%$.
Quantifying the latter of these is not straightforward, hence a pessimistic estimate has been given.

With these parameters Eq. (\ref{e2.10}) was evaluated using the algorithm described in section \ref{sec:alg}.
If the factor $\left( f(\epsilon_{\bf k}^n) - f(\epsilon_{{\bf k}'}^{n'}) \right)\tilde \delta'( \epsilon_{\bf k}^n-\epsilon_{{\bf k}'}^{n'} )$ in Eq. (\ref{e2.10}) was less than $10^{-3}$ the contribution to $\eta$ was not calculated, increasing the efficiency of the calculation.
Enough bands are included for the highest energy bands at each ${\bf k}$ point to be discarded.

Two parameters remain, which describe the temperature of the system, and how the discretisation of ${\bf k}$ space is dealt with through the smearing, $\Delta$.
The temperature enters through the Fermi-Dirac occupation functions in Eq. (\ref{e2.10}), and may be chosen to take any value.
For the metallic system considered here a weak temperature dependence is expected, so the zero-temperature limit is the quantity of interest.
We chose to use a `squashed Fermi-Dirac' distribution, $\tilde{f}(x)$, related to the function defined in Eq. (\ref{e2.7}) by
\begin{equation}
\tilde{f}(x) = 1-\int_{-\infty}^{x-\epsilon_{\mathrm{F}}} \tilde{\delta}(x) \mathrm{d}x.
\label{e4.1}
\end{equation}
If $\Delta_{\mathrm{T}}$ is the width associated with $\tilde{\delta}(x)$ then  the properties of the system are very close to those of a system described by a Fermi-Dirac distribution with a temperature given by $\Delta_{\mathrm{T}}/(\sqrt{2\pi} k_{\mathrm{B}})$, where $k_{\mathrm{B}}$ is the Boltzmann constant.
\begin{table*}
\label{defparagcl} 
\begin{center}
\begin{tabular}{r l c c}
\hline \hline
$T$                                              &
$\Delta_{\mathrm{T}}$                            & 
$\eta_{\uparrow} + \eta_{\downarrow}$            & 
$\eta_{\mathrm{\bf k}=\mathrm{\bf k}',\uparrow} + \eta_{\mathrm{\bf k}=\mathrm{\bf k}',\downarrow}$  \\
$(\mathrm{K})$ & 
$(\mathrm{eV})$ &
\multicolumn{2}{c}{$(\mathrm{meV}\ \mathrm{ps}\ $\AA$^{-\mathrm{2}})$} \\
\hline
  46   & 0.01 & 1.578 & 1.364 \\
 231   & 0.05 & 1.576 & 1.361 \\
 463   & 0.10 & 1.570 & 1.308 \\
 926   & 0.20 & 1.551 & 1.217 \\
1389   & 0.30 & 1.532 & 1.166 \\
1852   & 0.40 & 1.525 & 1.146 \\
2315   & 0.50 & 1.536 & 1.151 \\
\hline
\end{tabular}
\end{center}
\vspace*{.6cm}
\noindent
\caption{Variation of friction coefficient with temperature, $\Delta=0.5$ eV}
\label{tab1}
\end{table*}

Table \ref{tab1} shows results for $\eta$ summed over both spins, calculated over a range of temperatures for $\Delta=0.5$ eV (this is discussed below) and for both the isolated atom ($\eta_{\uparrow} + \eta_{\downarrow}$, see Eq. (\ref{e2.10})) and the coherent over-layer ($\eta_{\mathrm{\bf k}=\mathrm{\bf k}',\uparrow} + \eta_{\mathrm{\bf k}=\mathrm{\bf k}',\downarrow}$, see Eq. (\ref{e2.10b})).
Results for an isolated atom are weakly dependent on temperature, and we take $\Delta_{\mathrm{T}}=0.1$ eV to represent the low temperature limit.
For the coherent over-layer the temperature dependence is stronger, which is understandable since a greater proportion of the available (band to band) transitions are expected to be of higher energy.

The approximation that requires the most attention is the discretisation of ${\bf k}$ space and the reintroduction of a continuum through smearing.
As described in section \ref{sec:theory} we derive a continuous self-energy function using the $\tilde{\delta}(x)$ of Eq. (\ref{e2.7}) with width $\Delta$, and extract the linear behaviour of this function close to zero energy.
For a given set of ${\bf k}$ points we must find a value of $\Delta$ large enough for the approximate self-energy to have converged to a linear function near zero energy.
This critical $\Delta$ must be small enough (and so the density of the ${\bf k}$ point mesh must be high enough) to avoid the higher energy structure of the self-energy influencing the structure close to zero energy.

\begin{figure}
\begin{center}
\includegraphics*{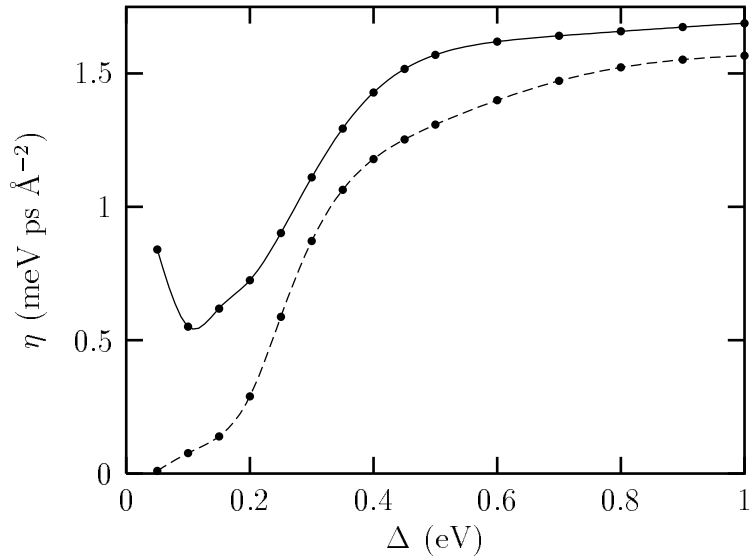}
\end{center}
\caption{Convergence of friction coefficient with the `smearing parameter' $\Delta$.
Results are given for $\Delta_{\mathrm{T}}=0.1$ eV ($T=463$ K).
The solid line shows the friction coefficient for an isolated atom above the surface, $\eta=\eta_{\uparrow} + \eta_{\downarrow}$, whereas the dashed line shows the friction coefficient for an over-layer of atoms in coherent motion, $\eta_{\mathrm{\bf k}=\mathrm{\bf k}',\uparrow} + \eta_{\mathrm{\bf k}=\mathrm{\bf k}',\downarrow}$.
}
\label{fig3}
\end{figure}

Results were calculated as above, with $\Delta_{\mathrm{T}}=0.1$ eV and $\Delta$ ranging from $0.05$ to $1.0$ eV.
These are shown in Fig.\ \ref{fig3} for both the isolated atom and coherent over-layer.
For the isolated atom $\eta$ shows good convergence by $\Delta=0.6$ eV, and this value is low enough to conserve the structure of the self-energy.
Convergence has not been achieved for the coherent over-layer, and the strong variation of the friction coefficient with $\Delta$ in this range suggests that the $54$ k-points are insufficient to achieve convergence for this system.
This difference in the convergence behaviour is due to far fewer transitions being available in the coherent over-layer system due to the $\mathbf{k}=\mathbf{k}'$ condition, and so fewer transitions to approximate a continuous self-energy.

\section{Conclusion}
\label{sec:conc}

An {\em ab initio\/} method has been presented that allows the evaluation of the friction due to electron-hole pair creation experienced by an isolated molecule in motion near a metal surface.
The approach described combines Kohn-Sham super-cell methods employing a plane-wave basis with a description of the electron-hole pair creation process via TDDFT.
Results have been presented for a H atom above a Cu(111) surface, and convergence of the calculations has been tested.
We find a significant difference in the results for motion of an isolated atom and those for a coherent over-layer of atoms, both in the physical properties of the system and in their numerical calculation.
Since the calculation is relatively expensive to perform for systems of interest an efficient parallel implementation of method has been given.

We wish to thank M. Persson and S. Holloway for useful discussions.


\begin{thebibliography}{}
\bibitem{bird97}      D.M. Bird, P.A. Gravil, {\em Surface Science\/} {\bf 377-379}, (1997) 555.
\bibitem{darling95}   G.R. Darling, S. Holloway, {\em Rep. Prog. Phys.\/} {\bf 58}, (1995) 1595.
\bibitem{gostein97}   M. Gostein, E. Watts, G.O. Sitz, {\em Phys. Rev. Lett.\/} {\bf 79}, (1997) 2891.
\bibitem{nienhaus99}  H. Nienhaus H, H.S. Bergh, B. Gergen, A. Majumdar, W.H. Weinberg, E.W. McFarland, {\em Phys. Rev. Lett.\/} {\bf 82}, (1999) 446.
\bibitem{huang00}     Y. Huang Y, A.M.  Wodtke, H. Hou, C.T. Rettner, D.J. Auerbach, {\em Phys. Rev. Lett.\/} {\bf 84}, (2000) 2985.
\bibitem{grimvall81}  G. Grimvall, {\em The Electron~Phonon Interaction in Metals\/}. North-Holland, Amsterdam, 1981.
\bibitem{hellsing84}  B. Hellsing, M. Persson, {\em Physica Scripta\/} {\bf 29} (1984) 360.
\bibitem{liebsch97}   A. Liebsch, {\em Phys. Rev.\/} {\bf B55} (1997) 13263.
\bibitem{headgordon92}M. Head-Gordon, J.C. Tully, {\em Phys. Rev.\/} {\bf B46} (1992) 1853.
\bibitem{amusia98}    M.Y. Amusia, V.R. Shaginyan, {\em Phys. Lett.\/} {\bf A250}, (1998) 157.
\bibitem{plihal98}    M. Plihal, D.C. Langreth, {\em Phys. Rev.\/} {\bf B58} (1998) 2191.
\bibitem{gross85}     E.K.U. Gross, W. Kohn, {\em Phys. Rev. Lett.\/} {\bf 55}, (1985) 2850.
\bibitem{inkson86}    J.C. Inkson, {\em Many-Body Theory of Solids\/} (Plenum Press, New York, 1986).
\bibitem{godby88}     R.W. Godby, M. Schluter, L.J. Sham, {\em Phys. Rev.\/} {\bf B37}, (1988) 10159.
\bibitem{white96}     J.A. White, D.M. Bird, M.C. Payne, {\em Phys. Rev.\/} {\bf B53}, (1996) 1667.
\bibitem{altmann91}   S.L. Altmann, {\em Band Theory of Solids: An Introduction from the Point of View of Symmetry\/} (Oxford University Press, Oxford, 1991).
\bibitem{gupta93}     A. Gupta, V. Kumar. {\em IEEE T. Parall. Distr.\/} {\bf 4} (1993) 922.
\bibitem{perdew96}    J.P. Perdew, K. Burke, M. Ernzerhof, {\em Phys. Rev. Lett.\/} {\bf 77}, (1996) 3865.
\bibitem{troullier91} N. Troullier, J. Martins, {\em Phys. Rev.\/}  {\bf B43},(1991) 1993.
\end{thebibliography}
\end{document}